\documentclass[traditabstract,rnote]{aa}

\usepackage{graphicx}
\usepackage{txfonts}
\usepackage{enumerate}

\begin{document} 

\title{The possibility of determining open-cluster
    parameters from BVRI photometry}
\subtitle{}

\author{H. Monteiro\thanks{E-mail: hektor.monteiro@gmail.com} and
  W. S. Dias} \institute{UNIFEI, DFQ - Instituto de Ci\^encias Exatas,
  Universidade Federal de Itajub\'a, Itajub\'a MG, Brazil}

\date{}

\begin{abstract}
  {In the last decades we witnessed an increase in studies of open
    clusters of the Galaxy, especially because of the good
    determination for a wide range of values of parameters such as
    age, distance, reddening, and proper motion. The reliable
    determination of the parameters strongly depends on the photometry
    available and especially on the U filter, which is used to obtain
    the color excess E(B-V) through the color-color diagram (U-B) by
    (B-V) by fitting a zero age main-sequence. Owing to the difficulty
    of performing photometry in the U band, many authors have tried to
    obtain E(B-V) without the filter. But because of the near
    linearity of the color-color diagrams that use the other bands,
    combined with the fact that most fitting procedures are highly
    subjective (many done ``by eye'') the reliability of those results
    has always been questioned. Our group has recently developed, a
    tool that performs isochrone fitting in open-cluster photometric
    data with a global optimization algorithm, which removes the need
    to visually perform the fits and thus removes most of the related
    subjectivity. Here we apply our method to a set of synthetic
    clusters and two observed open clusters (Trumpler 1 and Melotte
    105) using only photometry for the BVRI bands. Our results show
    that, considering the cluster structural variance caused only by
    photometric and Poisson sampling errors, our method is able to
    recover the synthetic cluster parameters with errors of less than
    10\% for a wide range of ages, distances, and reddening, which
    clearly demonstrates its potential. The results obtained for
    Trumpler 1 and Melotte 105 also agree well with previous
    literature values.}
\end{abstract}


\maketitle

\section{Introduction}

Our research group has a particular interest in using open clusters in
our studies of the spiral structure of our Galaxy. Some interesting
results were obtained with data of these objects that are compiled in
our catalog (\cite{Dias2002}): the rotation speed of the spiral
pattern as $25 \pm 1$ $km^{-1} s^{-1} kpc^{-1}$ and the co-rotation
radius at $1.06 \pm 0.08$ $\times$ the solar Galactocentric distance
(\cite{Dias2005}) and \cite{Lepine2008} determined the epicyclic
frequency to be $43 \pm 5$ $km^{-1} s^{-1} kpc^{-1}$ at the solar
radius.

In this kind of study and in many others that involve open clusters it
is crucial to accurately determine the fundamental parameters
(distance, reddening, and age) of the objects.  Traditionally this is
done using the color-magnitude diagrams via the main-sequence fitting
(MSF) technique, which in the majority of cases is fit to the data by
visual inspection. The work of \cite{PN06} (hereafter PN06) indicates
that the usual method applied up to now can determine distances of
open clusters with absolute errors of less than 20$\%$ (see all
details in the cited paper).

From the observational point of view, the determination of fundamental
parameters of open clusters requires accurate photometry (errors lower
than 0.03 in magnitude and lower than 0.05 in color index), deep
photometry (mainly for objects not previously studied) and also seems
to require data in the U-filter. The U-filter in principle allows a
better determination of the color excess and therefore leads to an
accurate determination of the distance modulus via MSF. The
determination of the color excess toward the cluster is performed
based on the observed photometric data, typically by visual
main-sequence fitting of the (B-V) vs. (U-B) diagram. Results on
determining the color excess without the use of the U-filter are
unreliable if done in the usual manner by visual fits because of the
near linearity of the color-color diagrams and the subjectivity of the
analysis.

We have addressed the subjectivity of the isochrone fits in
\cite{MDC10} (hereafter MDC10), where we developed a global
optimization tool using the cross-entropy optimization algorithm (CE)
to automatically fit open cluster UBV photometric data with well
defined criteria (see paper for details). We showed that the method
was robust and the results obtained for 10 open clusters agreed well
with previous studies found in the literature. Although the work used
only UBV photometry and the usual sequence to fit isochrones
(determine E(B-V) and then distance and age), the results obtained
with the global optimization indicated that it might be possible to
address the color excess determination without the use of the
U-filter. The advantages of such a procedure are clear, given that the
U-filter is possibly one of the most problematic ones to get good data
for, which typically requires longer exposure times.

In this work we investigate the possibility of determining
open cluster fundamental parameters from BVRI photometry alone. In
Sec. 2 we briefly describe the global optimization method developed in
MDC10 and the adaptations made for the present problem. In
Sec. 3 we describe benchmark tests using a set of synthetic open
clusters generated with known parameters. In Sec. 4 we apply the
method to the well studied open clusters Trumpler 1 and and Melotte
105 and discuss the results. In Sec. 5 we present our conclusions.

\section{The cross-entropy optimization method}

The CE procedure provides a simple adaptive way of estimating the
optimal model parameters. Basically, the CE method involves an
iterative procedure where each iteration consists of:

\begin{enumerate}
\item Random generation of the initial parameter sample, respecting
  pre-defined criteria;
\item Selection of the best candidates based on some mathematical
  criteria;
\item Random generation of updated parameter samples from the previous
  best candidates to be evaluated in the next iteration;
\item Optimization process repeats steps (ii) and (iii) until a
  pre-specified stopping criterion is fulfilled.
\end{enumerate}

In MDC10 we fully describe the algorithm and the fitting method
applied to open clusters and we refer the reader to that work and
references therein for more details. In short, the algorithm generates
a set of possible isochrone solutions (simulated open clusters) given
a pre-defined initial mass function, binary fraction, and
metallicity. Each generated solution is then compared to the observed
data through an objective function. The best solutions are selected, a
new set of solutions is generated from them and the process iterated
until convergence. In the present work we maintain the complete
structure of the previous algorithm discussed in detail in MDC10,
changing only the objective function to use the BVRI data instead of
only UBV.

In the case of isochrone fitting, the objective function used is the
likelihood of the data (in the present case BVRI photometry) for a
given model isochrone, which is obtained from the product of
probabilities of the $m$ stars observed and is given by:

\begin{eqnarray}
P(BVRI|{ISO}_{mod})= \sum_{m} \frac{1}{\sqrt{2\pi}\sigma_{B}\sigma_{V}\sigma_{R}\sigma_{I}} \times \nonumber\\ 
EXP-\frac{(B-B_{ISO})^2} {2\sigma_{B}} \times\nonumber\\
EXP-\left ( \frac{V-V_{ISO}} {2\sigma_{V}} \right )^2 \times \nonumber\\
EXP-\left ( \frac{R-R_{ISO}} {2\sigma_{R}} \right )^2 \times \nonumber\\
EXP-\left ( \frac{I-I_{ISO}} {2\sigma_{I}} \right )^2, 
\end{eqnarray}

where B, V, R, and I are the observed magnitudes and $\sigma_{B}$,
$\sigma_{V}$, $\sigma_{R}$, and $\sigma_{I}$ are the photometric
uncertainties in each filter. Here ${ISO}_{mod}$ designates the model
isochrone for a set of parameters (E(B-V), distance and age) and
$B_{ISO}$, $V_{ISO}$, $R_{ISO}$, and $I_{ISO}$ are the model
magnitudes for each point generated from the model.

\begin{equation}
\mathcal L = \prod P(BVRI|ISO_{mod}),
\end{equation}

where $\mathcal L$ is the likelihood obtained in the usual manner.  

As in MDC10, the tabulated isochrones are taken from \cite{GBBC00} and
\cite{MGB08} and are specified by four parameters, namely, age,
distance, extinction constant and metallicity. In the isochrone
fitting we defined the parameter space as follows:

\begin{enumerate}
\item {\bf Age}: from $log(age)=6.60$ to $log(age)=10.15$;
  \item {\bf distance}: from 1 to 10000 parsecs;
  \item ${\bf E(B-V)}$: from 0.0 to 3.0.
\end{enumerate}

To simplify the analysis we kept the metallicity constant at the solar
value. This should have no major impact on the evaluation of the
capability of the method to determine the fundamental parameters
because the changes due to metallicity are relatively small when
compared to the other parameters.

\section{Synthetic clusters}

\begin{figure}[!ht]
\begin{center}
\includegraphics[width=\columnwidth]{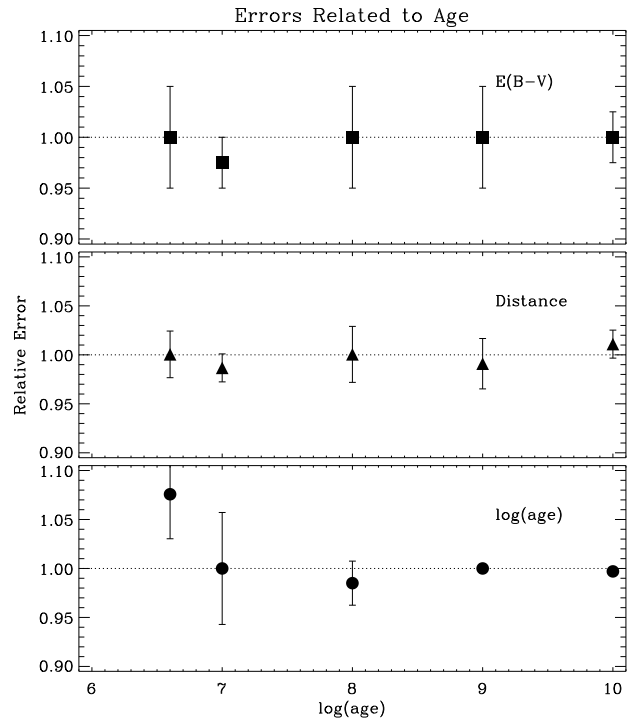}
\caption{ Relative errors for fits of the synthetic clusters fitted
  with our method and BVRI photometry only as a function of synthetic
  cluster age. The synthetic clusters have fixed $E(B-V)=0.4$,
  $3\sigma_{phot}=0.8\%$ and are at a distance of 1000~pc.}
\end{center}
\end{figure}  

\begin{figure}[!ht]
\begin{center}
\includegraphics[width=\columnwidth]{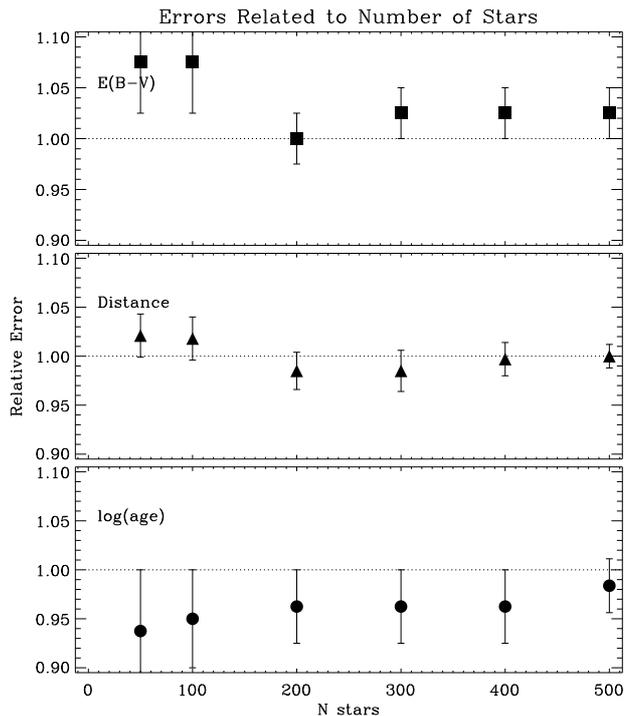}
\caption{ Relative errors for fits of the synthetic clusters fitted
  with our method and BVRI photometry only as a function of number of
  generated stars. The synthetic clusters have fixed $E(B-V)=0.4$,
  $3\sigma_{phot}=0.8\%$, $log(age)=8.0$ and are at a distance of
  1000~pc.}
\end{center}
\end{figure}  

To test the capability of recovering fundamental parameters using the
CE method we constructed a series of synthetic open clusters. The
generated clusters were chosen to be representative of clusters
present in the catalog (\cite{Dias2002}) and were generated from the
isochrone tables of \cite{GBBC00} and \cite{MGB08}, which are the same
as those used in the fitting algorithm.

We also generated clusters with distinct numbers of stars from low to
high density to evaluate the capacity of the method to retrieve the
parameters in unfavorable conditions.

All clusters were generated adopting solar metallicities, sampled from
a Salpeter IMF and with a 50\% binary population at a distance of
1000~pc, $E(B-V)=0.40$, $3\sigma_{phot}=0.8\%$ and limiting $m_V$
magnitude of 18. All fits were executed 20 times to perform an error
analisis via bootstrap re-sampling.

We did not include simulated contamination from field stars, but this
evaluation is already under way. But as we show in MDC10, the
filtering method we developed removed a good part of the contamination
and we expect that the main effect would likely be an increase in the
estimated uncertainty for the parameters.

The results obtained for the synthetic clusters fitted with our method
and BVRI photometry only are summarized in Fig.1 and Fig. 2 where the
relative errors for fits as a function of synthetic cluster age and
relative errors for fits as a function of number of generated stars
are shown. The distance and $E(B-V)$ were not varied in the results
presented in Figs. 1 and 2 because the major changes were caused by
the variation in age and number of stars. Varying the distance only
affects the photometric errors, which we assumed here to be some
percentage of the observed magnitudes, and so would only increase the
final uncertainty. In Fig. 2 the age was also fixed at $log(age)=8.0$.

To check the reliability of the method under distinct reddening
values, we performed a set of tests with synthetic clusters generated
with $E(B-V)$ varying from 0.3 to 3.0, covering the observed parameter
range, and with $log(age)=8.70~yr$ at a distance of 2100 pc. We note
that an increase in the magnitude limit to $m_V=22$ was necessary for
the synthetic clusters with $E(B-V) > 2$ so that the main sequence
could be minimally sampled. The results presented in Fig. 3 indicate
that the uncertainty of the distance determination tends to increase
with increasing $E(B-V)$, which is to be expected. Interestingly, the
uncertainties obtained for the various reddening values studied were
essentially constant around 0.03, which shows up as declining relative
errors in Fig.2.

\begin{figure}[!ht]
\begin{center}
\includegraphics[width=\columnwidth]{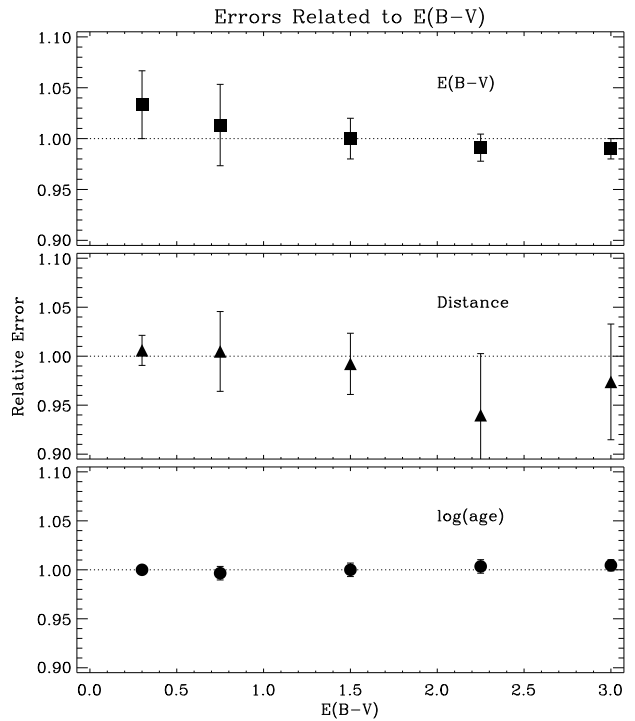}
\caption{ Relative errors for fits of the synthetic clusters fitted
  with our method and BVRI photometry only, as a function of number of
  generated $E(B-V)$ values. The synthetic clusters have fixed
  $log(age)=8.70~yr$ at a distance of 2100 pc.}
\end{center}
\end{figure}

We also tested the fitting method on a synthetic cluster generated
with the same parameters as the one presented in Sec. 5.1 of MDC10,
i.e., 400 stars with photometric error $3\sigma_{phot}=1\%$ sampled
from an isochrone with $log(age)=8.70~yr$, at a distance of 2100 pc,
$E(B-V)$=0.40 and Z=0.019. The results obtained from the fitting
algorithm were

\begin{itemize}
\item $log(age)=(8.73 \pm 0.05)~yr$
\item $distance=(2100 \pm 70)~pc$
\item $E(B-V)=(0.39 \pm 0.03)$
\end{itemize}

The results indicate that the method using only BVRI is capable of
recovering the cluster parameters with about the same accuracy as the
UBV method presented in MDC10. The precision is also about the same
exept for the reddening, where the BVRI-fitted value shows larger
uncertainty.

\section{Application to observed clusters}

We applied this new approach of fitting isochrones to a couple of
well-studied open clusters. The selected clusters were Trumpler 1 and
Melotte 105, which have observational data in the BVRI filters obtained
by \cite{YS02} (hereafter YS02) and \cite{SMB01} (hereafter SMB01)
respectively. The main reason for choosing these particular clusters
is that they are the only ones that have BVRI data and were also
studied in our previous work (MDC10), where we applied the CE method
using UBV photometry only, following the traditional route of first
determining the reddening through the (U-B) versus (B-V) diagram and
then fitting the theoretical isochrones.

The results obtained for Trumpler 1 and Melotte 105 using our fitting
method and the BVRI photometry agrees very well with the results
obtained by YS02 and SMB01 and the results from our previous work
using only UBV photometry. The values also agree with those of
PN06. In Fig.~4 we show the final-fit results (isochrones) for
Trumpler 1 and in Fig.~5 for Melotte 105, overplotted on
color-magnitude diagrams where the symbol sizes are proportional to
the weight determined for each star based on the filtering scheme
described in MDC10.

For Trumpler 1 the results obtained from our BVRI fit agrees well with
the values obtained by YS02 (see Table 1). Some small but significant
differences are found when we consider the results of PN06 and MDC10,
but these are likely due to weights given to the bright stars as
discussed in MDC10. The weight of the bright stars are likely the main
cause for the age differences. Notice, however, that using the BVRI
only we obtain almost the same E(B-V) as in the fit made with the
(U-B) versus (B-V) diagram.

For Melotte 105 we find a similar situation where we found small
significant differences in E(B-V) and distance when comparing our
results with those of SMB01. The differences are again likely due to
distinct weights given to particular stars. The weight might be
particularly relevant when comparing the results of MDC10, where the
red giants received a lower weight compared to this work, which could
be the reason for the lower distance encountered in that work.

\begin{table}
\caption{CE fit and literature results for Trumpler 1}
\label{table:1}      
\centering                          
\begin{tabular}{c c c c}        
\hline\hline                 
   Ref       & $E(B-V)$        & Distance (pc) & log(Age(yr)) \\    
\hline                        
   PN06      & $0.57 \pm 0.04$ & $2356\pm 511$ & $7.5 \pm 0.1$ \\
   MDC10     & $0.59 \pm 0.05$ & $2419\pm 185$ & $7.8 \pm 0.2$ \\
   YR02      & $0.60 \pm 0.05$ & $2600\pm 100$ & $7.6 \pm 0.1$ \\
   this work & $0.60 \pm 0.03$ & $2706\pm 120$ & $7.2 \pm 0.3$ \\
\hline                                
\end{tabular}
\end{table}

\begin{table}
\caption{CE fit and literature results for Melotte 105}
\label{table:1}      
\centering                          
\begin{tabular}{c c c c}        
\hline\hline                 
   Ref       & $E(B-V)$        & Distance (pc) & log(Age(yr)) \\    
\hline                        
   PN06      & $0.48 \pm 0.05$ & $2094\pm 159$ & $8.4 \pm 0.1$ \\
   MDC10     & $0.47 \pm 0.02$ & $1750\pm 111$ & $8.4 \pm 0.1$ \\
   SMB01     & $0.52 \pm 0.05$ & $2300\pm 200$ & $8.4 \pm 0.1$ \\
   this work & $0.42 \pm 0.03$ & $1932\pm 95$  & $8.5 \pm 0.1$ \\
\hline                                
\end{tabular}
\end{table}

\begin{figure*}[!h]
\begin{center}
\includegraphics[scale=0.55]{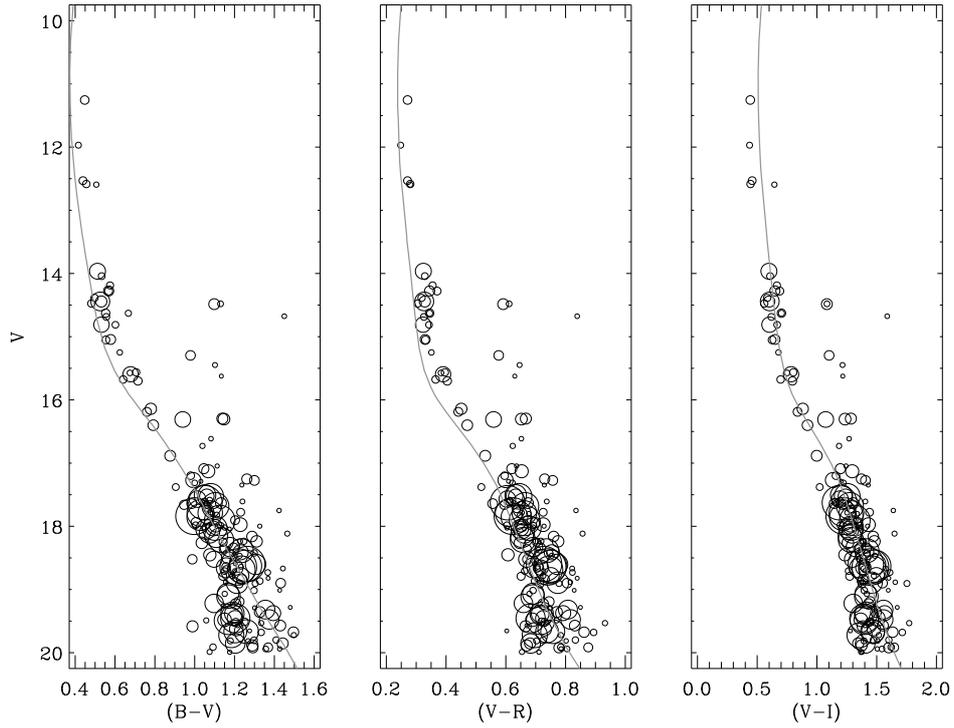}
\caption{ Final-fit results for Trumpler 1 overplotted on
  color-magnitude diagrams where the symbol sizes are proportional to
  the weight determined for each star as described in MDC10.}
\end{center}
\end{figure*}  

\begin{figure*}[!h]
\begin{center}
\includegraphics[scale=0.55]{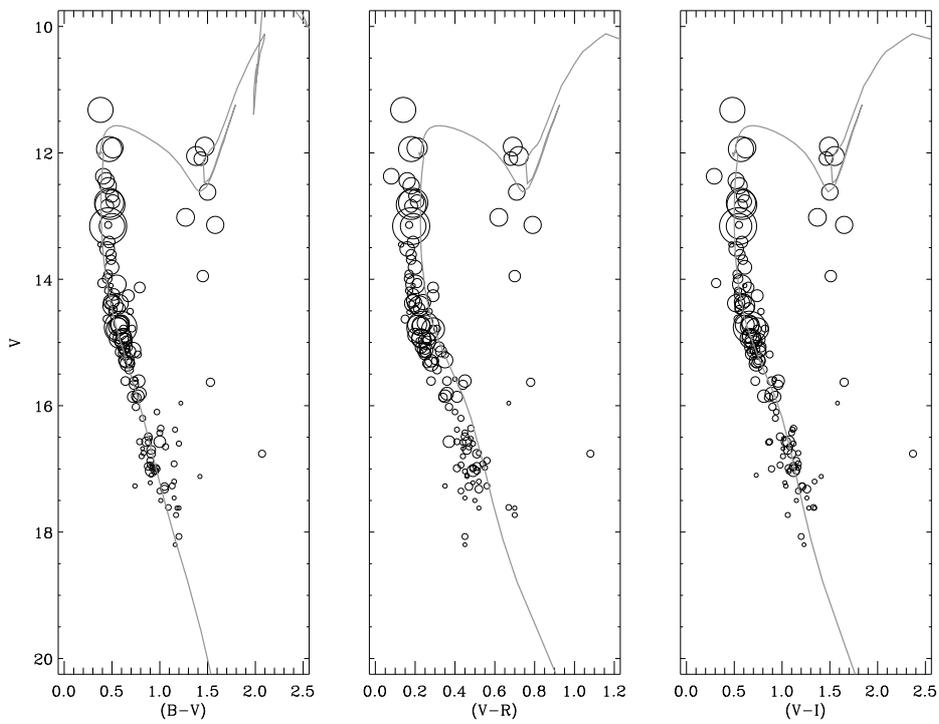}
\caption{Same as Fig. 3 for Melotte 105.}
\end{center}
\end{figure*}

\section{Conclusions}

We investigated the possibility of fitting theoretical tabulated
isochrones to open cluster BVRI photometry with a global optimization
method developed in MDC10. In the context of upcoming ground-based
telescopes and space missions that will produce massive amounts of
data, this tool, wich additionally has the possibility of full
automation, is of great value.

The results obtained with synthetic open clusters that are generated
with parameters that cover typical values found in the literature show
that even without the use of the U-filter, fundamental parameters can
be determined with a relative precision of about 10\% or better in
most cases in the studied parameter space, also with an accuracy
similar to the method described in MDC10 using UBV photometry only.
In particular we showed that the reddening can be determined for a wide
range of cluster ages, despite the lack of visual structure in the
color-color diagrams of the clusters. Even when considering objetcs
with few stars the method was able to recover the original parameters
within the uncertainties, showing some systematic behavior for
clusters with a low number of stars as shown in Fig. 2.

It is important to point out that the relative uncertainties discussed
previously are to be taken as lower limits to the real expected
uncertainties that are attainable with real data. The errors reported
are only due to the adopted error in the photometry and Poisson
statistics that arise from drawing a finite number of stars from model
isochrones. Real clusters will have structural variations because of
binary fraction or chemical composition among many other factors, will
affect the final uncertainty of the method. Because it is difficult to
quantify the nature of these factors and their effect on the final
uncertainty, a definitive error analysis should be carried out in a
case-by-case basis.

Applying the method to two observed clusters also studied with our
optimization method and UBV photometry (see MDC10 for details) showed
that the use of BVRI filters only also produced consistent
results. The results agree well with those obtained by previous works,
in particular when considering the results of PN06, where the authors
included a large number of independent literature
determinations. Assuming that the results of PN06 are the best
estimates of the real cluster parameters, we can estimate a better
value of the final uncertainty of the method for these objetcs. For
Trumpler 1 the highest relative uncertainty comes from the distance
and is about 15\%. For Mellote 105 the largest relative error is due
to the reddening being approximately 13\%. These values are likely
more realistic estimates of the real uncertainty of the method for
these two clusters.

Our results show that it is indeed possible to reliably determine
open-cluster fundamental parameters from BVRI photometry only despite
the lack of visual features in the color-color diagrams.

\section{Acknowledgements} 

H. Monteiro would like to thank CNPq (grant number
470135/2010-7). W. S. Dias thanks CNPq (grant number 302762/2007-8)
and FAPEMIG (process number APQ-00090-08). We also thank the referee
Dr. Bailer-Jones for extremely useful comments that greatly improved
the paper.


\end{document}